\documentclass[12pt]{article}

\usepackage[T1]{fontenc}
\usepackage{amsmath,amssymb}
\usepackage{graphicx}
\usepackage{geometry}
\usepackage{bm}

\newcommand{\rhat}{\hat{r}}
\newcommand{\vecgamma}{\vec{\gamma}}

\begin{document}

\begin{center}
{\large\bfseries Influence of the Electron's Anomalous Magnetic Dipole Moment
on High-Atomic-Number Atoms}\\[1.2em]
{C. A. S. Almeida$^{1,2}$ \quad and \quad J. Auto-Neto$^{3}$}\\[1em]
{\small
$^{1}$ Centro Brasileiro de Pesquisas F\'isicas -- CBPF/CNPq,
Rua Dr.\ Xavier Sigaud 150, 22290 -- Rio de Janeiro, RJ -- Brazil\\[0.3em]
$^{2}$ On leave of absence from Departamento de F\'isica da
Universidade Federal do Cear\'a\\[0.3em]
$^{3}$ Departamento de F\'isica da Universidade Federal do Cear\'a,
Caixa Postal 6030, 60000 -- Fortaleza, CE -- Brazil}\\[1.5em]
{\small CBPF-NF-027/90 \qquad Rio de Janeiro, 1990}
\end{center}

\vspace{0.5em}
\noindent\rule{\textwidth}{0.4pt}
\begin{quote}\small
\textbf{Transcription note (2026).} This is a faithful LaTeX transcription of the
preprint CBPF-NF-027/90 (Centro Brasileiro de Pesquisas F\'isicas, Rio de Janeiro,
1990), which was never published in a journal. The text, equations, and the two
figures reproduce the original document; only the typesetting has been modernized
and a small number of obvious typographical slips in the original have been left
as they were, to preserve fidelity. The figures are cleaned scans of the originals.
J.\ Auto-Neto, the corresponding author's M.Sc.\ advisor, passed away in 2015; this
transcription is deposited in his memory.
\end{quote}
\noindent\rule{\textwidth}{0.4pt}
\vspace{1em}

\begin{abstract}
Super-heavy atoms ($Z > 100$) are usually studied in the context of the so-called
``Quantum Electrodynamics of Strong Fields''. In this theory the problem of the
singularity in the electron energy whenever $Z > 137$ is overcome. This is done by
considering the finite size of the nucleus and leads to interesting phenomena, such
as the spontaneous production of positrons. Here, we show that taking into account
the contribution from the Anomalous Magnetic Dipole Moment of the electron (by means
of an effective theory), within a point-nucleus model, is a sufficient condition to
obtain regular wave functions and physically acceptable energy values for $Z > 137$.

\end{abstract}

\section{Introduction}

In the last fifteen years a great deal of attention has been given to the problem of
atoms with very high atomic numbers, $Z > 137$. At sufficiently small distances the
radial solutions of the Dirac equation for an electron in the field of a point charge
$Z|e|$ behave as $r^{\pm\gamma}$, where
\begin{equation}
\gamma = \left[\,(j + \tfrac{1}{2})^2 - (Z\alpha)^2\,\right]^{1/2}
\label{eq:gamma}
\end{equation}
and $\alpha = e^2 \simeq 1/137$ is the fine structure constant. Now if
$Z\alpha > j + \tfrac{1}{2}$, $\gamma$ becomes a pure imaginary number and the solution
oscillates infinitely many times as $r \to 0$ and no boundary condition can be imposed
at $r = 0$.

This difficulty disappears when one acknowledges the finite size of the nucleus. This
provides a cut-off for the attractive Coulomb field and meaningful solutions which
behave regularly at the origin can be found for any $Z$. The theory presents some
interesting novel features such as the decay of the neutral vacuum into a charged
vacuum by spontaneous emission of positrons [1,2,3,4]. These predictions have been
observed in U--Cm and U--U collisions [5].

In this paper we present an alternative way to regularize the wave-functions at
$r = 0$. As Grandy [6] pointed out, this can be done if one takes into account the
anomalous magnetic dipole moment (AMDM) of the electron through the Pauli effective
Hamiltonian (see below). We then show that well behaved functions can be obtained,
even in the case of a point nucleus. The system behaves pretty much as the
finite-nucleus model and all the interesting new phenomena which appear therein are
also present in this new focalization.

\section{Anomalously magnetized electron in the field of a point nucleus}

The AMDM of the electron can be taken into account by adding to Dirac's equation a
phenomenological new term due to Pauli. The thus modified equation is [7]
\begin{equation}
\left[\,\gamma^\mu (p_\mu - e A_\mu) - m\,\right]\psi
= \left(\frac{ae}{2m}\right)\frac{1}{2}\,\sigma^{\mu\nu} F_{\mu\nu}\,\psi
\label{eq:pauli}
\end{equation}
Here, $a = (g-2)/2$ is the ratio between the AMDM and the Bohr magneton. We shall take
a static point nucleus with charge $Z|e|$, so that the potential 4-vector is given by
\begin{equation}
A^\mu = \left(Z|e|/r,\ \vec{0}\,\right)
\label{eq:Amu}
\end{equation}
Then \eqref{eq:pauli} can be put in the form
\begin{equation}
i\,\frac{\partial\psi}{\partial t} = H\psi
\label{eq:schro}
\end{equation}
with
\begin{equation}
H = \gamma^0\,\vecgamma\cdot\vec{p} + m\gamma^0
- \frac{Z\alpha}{r}
- \left(\frac{iaZ\alpha}{2m r^2}\right)\vecgamma\cdot\rhat
\label{eq:H}
\end{equation}
where $\rhat = \vec{r}/|\vec{r}|$ is the unit radial vector and $\gamma^0$ and
$\vecgamma$ are the Dirac matrices. Now, since the anomalous term
$(-iaZ/2mr^2)(\vecgamma\cdot\rhat)$ commutes with $J^2$, $J_z$ and with the parity
operator, the wave function can be put in the form
\begin{equation}
\psi = \frac{1}{r}
\begin{pmatrix}
U_1(r)\,\Omega^m_k(\rhat)\\[0.6em]
i\,U_2(r)\,\Omega^m_k(\rhat)
\end{pmatrix}
\label{eq:psi}
\end{equation}
where $\Omega^m_k$ is a spherical Pauli spinor and $-k = \pm(j + \tfrac{1}{2})$ is the
eigenvalue of the spin-orbit operator $\gamma^0(\vec{\Sigma}\cdot\vec{L}+1)$. The radial
functions $U_1$ and $U_2$ satisfy the coupled equations [8]
\begin{align}
\frac{dU_1}{dr} + \left(\frac{k}{r} - \frac{aZ\alpha}{2m r^2}\right)U_1
&= \left(m + E + \frac{Z\alpha}{r}\right)U_2,
\nonumber\\[0.4em]
\frac{dU_2}{dr} - \left(\frac{k}{r} - \frac{aZ\alpha}{2m r^2}\right)U_2
&= -\left(m - E - \frac{Z\alpha}{r}\right)U_1.
\label{eq:coupled}
\end{align}

\section{The wave-function for \boldmath$r \ll 1/m$}

The set of equations \eqref{eq:coupled} has two essential singularities at $r = 0$ and
$r = \infty$ and is very difficult to solve exactly. Nevertheless, its behaviour near
the origin can be found by neglecting the terms $m + E$ and $m - E$ which, in this
region, are small compared to $Z\alpha/r$. By introducing $x = mr$, we then have the
truncated equations
\begin{align}
\frac{dU_1}{dx} + \left(\frac{k}{x} - \frac{aZ\alpha}{2x^2}\right)U_1
&= \frac{Z\alpha}{x}\,U_2,
\nonumber\\[0.4em]
\frac{dU_2}{dx} - \left(\frac{k}{x} - \frac{aZ\alpha}{2x^2}\right)U_2
&= -\frac{Z\alpha}{x}\,U_1,
\label{eq:trunc}
\end{align}
which are freed of the essential singularity at $x = \infty$. The decoupling of these
equations leads to
\begin{align}
\frac{d^2U_1}{dx^2} + \frac{1}{x}\frac{dU_1}{dx}
+ \left[\frac{(Z\alpha)^2 - k^2}{x^2}
+ \frac{(2k+1)aZ\alpha}{2x^3}
- \frac{(aZ\alpha)^2}{4x^4}\right]U_1 &= 0,
\nonumber\\[0.4em]
\frac{d^2U_2}{dx^2} + \frac{1}{x}\frac{dU_2}{dx}
+ \left[\frac{(Z\alpha)^2 - k^2}{x^2}
+ \frac{(2k-1)aZ\alpha}{2x^3}
- \frac{(aZ\alpha)^2}{4x^4}\right]U_2 &= 0.
\label{eq:decoupled}
\end{align}
Because $x = \infty$ is a regular point we introduce a new variable
\begin{equation}
y = aZ\alpha/x
\label{eq:y}
\end{equation}
and new functions
\begin{equation}
v_{1,2} = y^{1/2}\,U_{1,2}
\label{eq:v}
\end{equation}
in terms of which, \eqref{eq:decoupled} becomes
\begin{align}
\frac{d^2v_1}{dy^2} + \left[-\frac{1}{4} + \frac{k+1/2}{y}
+ \frac{(Z\alpha)^2 - k^2 + 1/4}{y^2}\right]v_1 &= 0,
\nonumber\\[0.4em]
\frac{d^2v_2}{dy^2} + \left[-\frac{1}{4} + \frac{k-1/2}{y}
+ \frac{(Z\alpha)^2 - k^2 + 1/4}{y^2}\right]v_2 &= 0.
\label{eq:whitteq}
\end{align}
The solutions of \eqref{eq:whitteq} which satisfy the boundary conditions
$U_1(x),\,U_2(x) \to 0$ as $x \to 0$ ($y \to \infty$) are Whittaker functions of
second kind [9]
\begin{align}
v_1 &= A_1\,W_{k+1/2,\,\gamma}(y),
\nonumber\\
v_2 &= A_2\,W_{k-1/2,\,\gamma}(y),
\label{eq:whitt}
\end{align}
where $\gamma$ is given in \eqref{eq:gamma}. To determine the relation between $A_1$
and $A_2$ we substitute \eqref{eq:whitt} back in \eqref{eq:trunc} thus obtaining
\begin{equation}
A_2/A_1 = -Z\alpha.
\label{eq:ratio}
\end{equation}
So that, considering \eqref{eq:y} and \eqref{eq:v}, we have
\begin{align}
U_1 &= A\,(x/aZ\alpha)^{1/2}\,W_{k+1/2,\,\gamma}(aZ\alpha/x),
\nonumber\\
U_2 &= -A Z\alpha\,(x/aZ\alpha)^{1/2}\,W_{k-1/2,\,\gamma}(aZ\alpha/x),
\label{eq:Usol}
\end{align}
where $A$ is a normalizing constant. The Whittaker function remains real as $Z\alpha$
surpasses $j + 1/2$ and $\gamma$ becomes imaginary [3,4] so no problem appears in the
boundary condition at $r = 0$. From the asymptotic behaviour of the Whittaker function
\begin{equation}
W_{m,\gamma}(y) \sim y^m\,e^{-y/2} \quad\text{as}\quad y \to \infty
\label{eq:asympt}
\end{equation}
we obtain the behaviour of the wave function near $r = 0$,
\begin{align}
U_1(x) &\simeq A\,(aZ\alpha/x)^k\,e^{-aZ\alpha/2x},
\nonumber\\
U_2(x) &\simeq A Z\alpha\,(aZ\alpha/x)^{k-1}\,e^{-aZ\alpha/2x},
\qquad\text{as } x \to 0.
\label{eq:nearzero}
\end{align}
From \eqref{eq:nearzero} we see that, as long as $a > 0$, the exponential factor
dominates and the wave function dies off as $x \to 0$. (If $a < 0$, one makes the
substitution $y = -aZ\alpha/x$ instead of \eqref{eq:y} and again, one gets well behaved
solutions near $x = 0$ which can be obtained from \eqref{eq:Usol} by the transformation
$a \to -a$, $k + 1/2 \to -k \mp 1/2$.) Note that near the origin the wave function is
independent of $\gamma$. The effect that the AMDM can render an ill-defined Dirac
equation self-adjoint was noticed before by Goldhaber et al.\ [10] in the context of
the Dirac equation in the field of a magnetic monopole.

\section{Numerical results and qualitative analysis}

In order to find the eigenvalues of energy we decoupled the set of equations
\eqref{eq:coupled} and made the transformation [8]
\begin{equation}
U_i = \left[\,m + (E + Z\alpha/r)\,\varepsilon_i\,\right]^{1/2}\chi_i(r)
\label{eq:Uchi}
\end{equation}
where $i = 1,2$; $\varepsilon_1 = +1$; $\varepsilon_2 = -1$. By defining
\begin{equation}
h_i(r) = 1 + (E - m\varepsilon_i)\,r/Z\alpha
\label{eq:hi}
\end{equation}
we find that the transformed functions $\chi_i$ satisfy the Schr\"odinger-type equation
\begin{equation}
-\chi_i''(r) + V_{\text{eff}}^{(i)}(r)\,\chi_i(r) = \omega^2\,\chi_i(r)
\label{eq:schrtype}
\end{equation}
where
\begin{equation}
\omega^2 = E^2 - m^2
\label{eq:omega}
\end{equation}
and
\begin{align}
V_{\text{eff}}^{i} &= -2EZ\alpha/r
+ \left[\,k(k + \varepsilon_i) - (Z\alpha)^2\,\right]/r^2
- (1 + k\varepsilon_i)/r^2 h_i(r)
\nonumber\\
&\quad
+ 3/4 r^2 h_i^2(r)
- (k + \varepsilon_i)\,aZ\alpha\varepsilon_i/m r^3
+ aZ\alpha\varepsilon_i/2m r^3 h_i(r)
+ (aZ\alpha)^2/4m^2 r^4.
\label{eq:Veff}
\end{align}
Observe that the energy-dependent effective potential \eqref{eq:Veff} has a repulsive
term $(aZ\alpha)^2/4m^2 r^4$ which at small distances dominates and cuts off the
Coulomb attraction. This $1/r^4$ term plays here a role analogous to the cut-off in
the potential of the finite-nucleus model.

In order to solve \eqref{eq:schrtype} numerically, one first has to decide which value
of $a$ to use in the effective potential \eqref{eq:Veff}. If we take the infrared limit
$a = \alpha/2\pi$, then \eqref{eq:nearzero} indicates the AMDM leads to a cut-off of the
wave function at distances $aZ\alpha/2m r \sim 1$ or $r \sim aZ\alpha/4\pi m < 1$~fm,
corresponding to momenta of almost 1~GeV/c. However, the infrared limit of the AMDM is
obtained from the form factor in the limit $|\vec{p}\,| \ll m \sim 0.5$~MeV/c. It is
therefore questionable if the infrared limit can be applied. On the other hand Ritus
[11] has shown that in the context of a uniform electric field $\epsilon$ (as well as
magnetic) the AMDM depends on the field and on the momentum component transverse to the
field. For null transverse momentum and very strong field ($e\epsilon/m^2 \gg 1$) the
AMDM is twice that of the infrared limit, that is $a = \alpha/\pi$. In the case
considered here, the electron gets so close to the nucleus that the above condition on
the field is certainly satisfied. So, it is reasonable to guess that, at least for S
waves (which have no transverse momentum to the field) the same limit may be applied.

Equation \eqref{eq:schrtype} was solved numerically in both limits. The eigenvalues
thus obtained are plotted in Fig.~1 (together with the finite-nucleus model results
[12] for comparison). The numerical method we used requires that
$(\partial V_{\text{eff}}/\partial E) - 2E$ has the same sign for all $r$. This
condition is satisfied only for $i = 1$, $k < 0$. We were thus unable to verify if the
fast diving of the $2P_{1/2}$ state ($k = +1$) observed for extended nuclei also happens
in the case studied here. The shape of the curves in Fig.~1 can be qualitatively
understood as follows: suppose we know the eigenvalues of the Hamiltonian \eqref{eq:H}
at a given $Z$ value
\begin{equation}
H(Z)\,\psi_{nk} = E_{nk}(Z)\,\psi_{nk}
\label{eq:HZ}
\end{equation}
and let us examine what happens when the atomic number is increased by a small amount
$\delta Z$. We have
\begin{equation}
H(Z + \delta Z) = H(Z) + \delta Z\,U
\label{eq:HdZ}
\end{equation}
where
\begin{equation}
U = -\alpha/x - (ia\alpha/2x^2)\,\vecgamma\cdot\rhat .
\label{eq:U}
\end{equation}
The variation in energy can be found by perturbation theory. In first order we have
\begin{equation}
\delta E = \delta Z\,\langle\psi|U|\psi\rangle
\label{eq:dE}
\end{equation}
then, using \eqref{eq:psi} we get
\begin{equation}
\frac{dE}{dZ} = -\alpha\int dx\,(U_1^2 + U_2^2)/x
- a\alpha\int dx\,U_1 U_2/x^2.
\label{eq:dEdZ}
\end{equation}
In the case $a = 0$, \eqref{eq:dEdZ} looses its meaning for $Z\alpha > 1$, because the
integrals diverge. For $a > 0$, however, the exponential factors in \eqref{eq:nearzero}
ensure the convergence for any $Z$, as long as $|E| < m$. The first term in
\eqref{eq:dEdZ} is negative. The second is positive, at least in the ground state,
since, in this case, $U_1$ and $U_2$ have opposite signs, as one can see from
\eqref{eq:nearzero} and from the fact that the ground state wave function has no nodes.
Nevertheless, since $a$ is so small ($\sim 10^{-3}$), the second term contributes little
and $dE/dZ < 0$.

Furthermore, one can see that $d^2E/dZ^2$ is also negative. The most important
contribution to the integrals in \eqref{eq:dEdZ} comes from the region around the
maxima of $U_1$ and $U_2$. As $Z$ increases, these maxima move towards $x = 0$ (the
electron gets closer to the point nucleus), therefore, increasing the contribution of
the $1/x$ factor in the first integral. So, we can presume that the derivative $dE/dZ$
increases in absolute value as $Z$ increases. So, $E$ is a decreasing function of $Z$
with negative curvature.

It is then clear that, just as in the finite-nucleus model, as the atomic number
increases, a value, called critical $Z$, will be attained, for which $E = -m$. Evidently
this number will depend on the quantum numbers $n$ and $k$ of \eqref{eq:HZ}. The largest
integer for which $E > -m$ was found to be $Z = 159$ (198) for 1S (2S) waves and
$a = \alpha/2\pi$. For $a = \alpha/\pi$ these values change to 164 and 209, respectively.
As $Z$ increases further, the previously discrete state will become a member of the
negative continuum as \eqref{eq:dEdZ} indicates. These over-critical states can be
singled out from other members of the continuum by the shape of their wave-functions.
The wave function of the continuum states are practically zero within the atom but, as
the energy approaches the ``dived in'' state energy, a highly pronounced peak appears
near $r = 0$, which is characteristic of resonant states. This situation is illustrated
in Fig.~2, for $a = \alpha/2\pi$. To obtain this ``dived in'' resonant 1S state we
extrapolated the $E(Z)$ function obtained for the undercritical 1S state to $Z = 160$,
the first over-critical atomic number. Next we put this energy value back in
\eqref{eq:coupled} and solved the resulting set of equations by an initial value method.
The initial values for $U_1$ and $U_2$ were obtained from \eqref{eq:Usol} at
$x = mr = 10^{-5}$. Then by a careful search around this energy value we looked for the
resonant behaviour described above and found an extremely narrow peak at
$E = -1.074969485$ electron masses. Fig.~1 also shows the wave function for a non
resonant energy value.

\begin{figure}[p]
\centering
\includegraphics[width=0.92\textwidth]{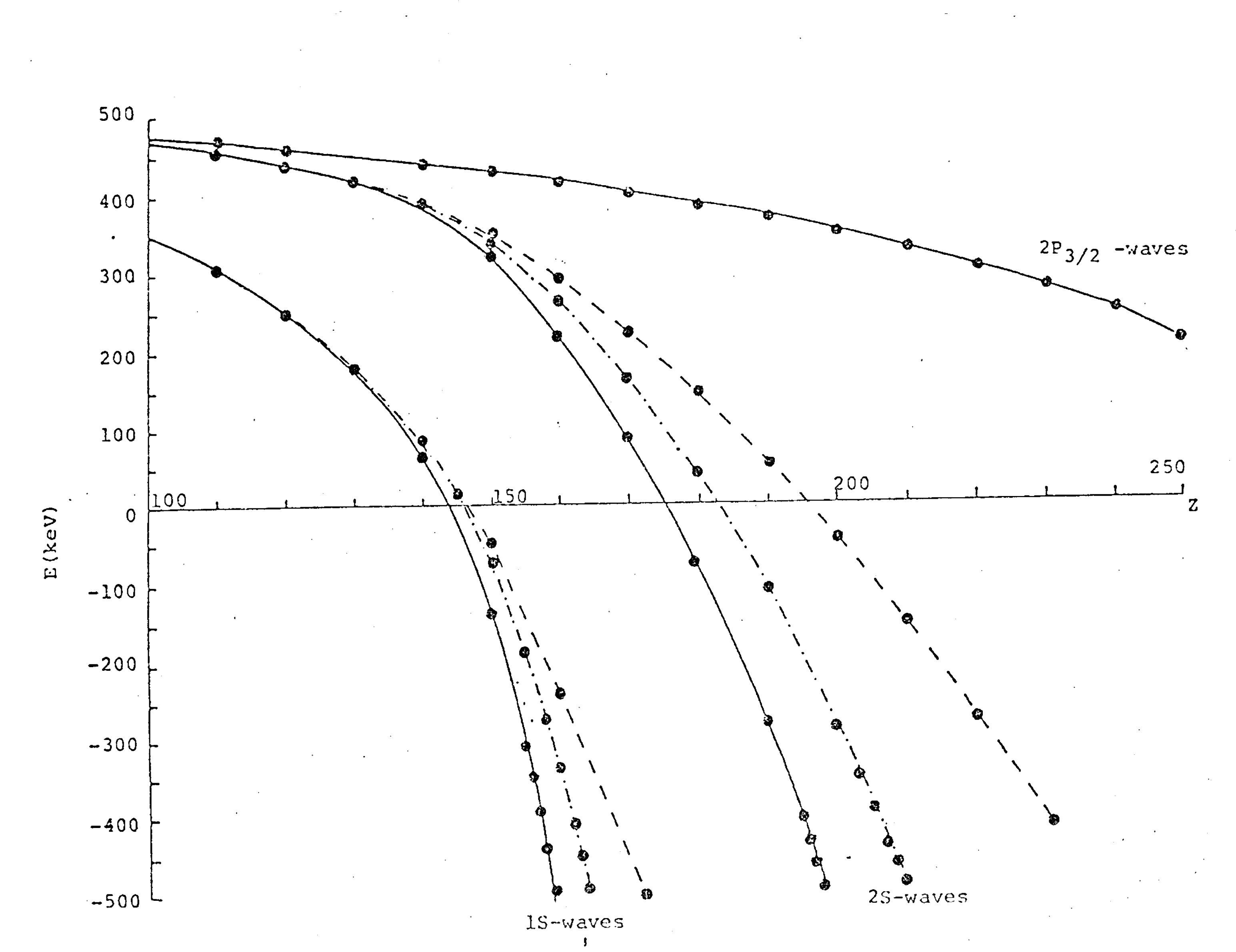}
\caption{$E\times Z$ curves.
\textemdash\,\textemdash\,\textemdash\ finite nucleus [12];
\rule[0.4ex]{1.2em}{0.4pt}\ point nucleus, AMDM model, $a = \alpha/2\pi$;
\textendash$\,\cdot\,$\textendash$\,\cdot\,$\textendash\ point nucleus, AMDM model,
$a = \alpha/\pi$. (For $2P_{3/2}$ waves, the three curves coincide.)}
\label{fig:1}
\end{figure}

\begin{figure}[p]
\centering
\includegraphics[width=0.78\textwidth]{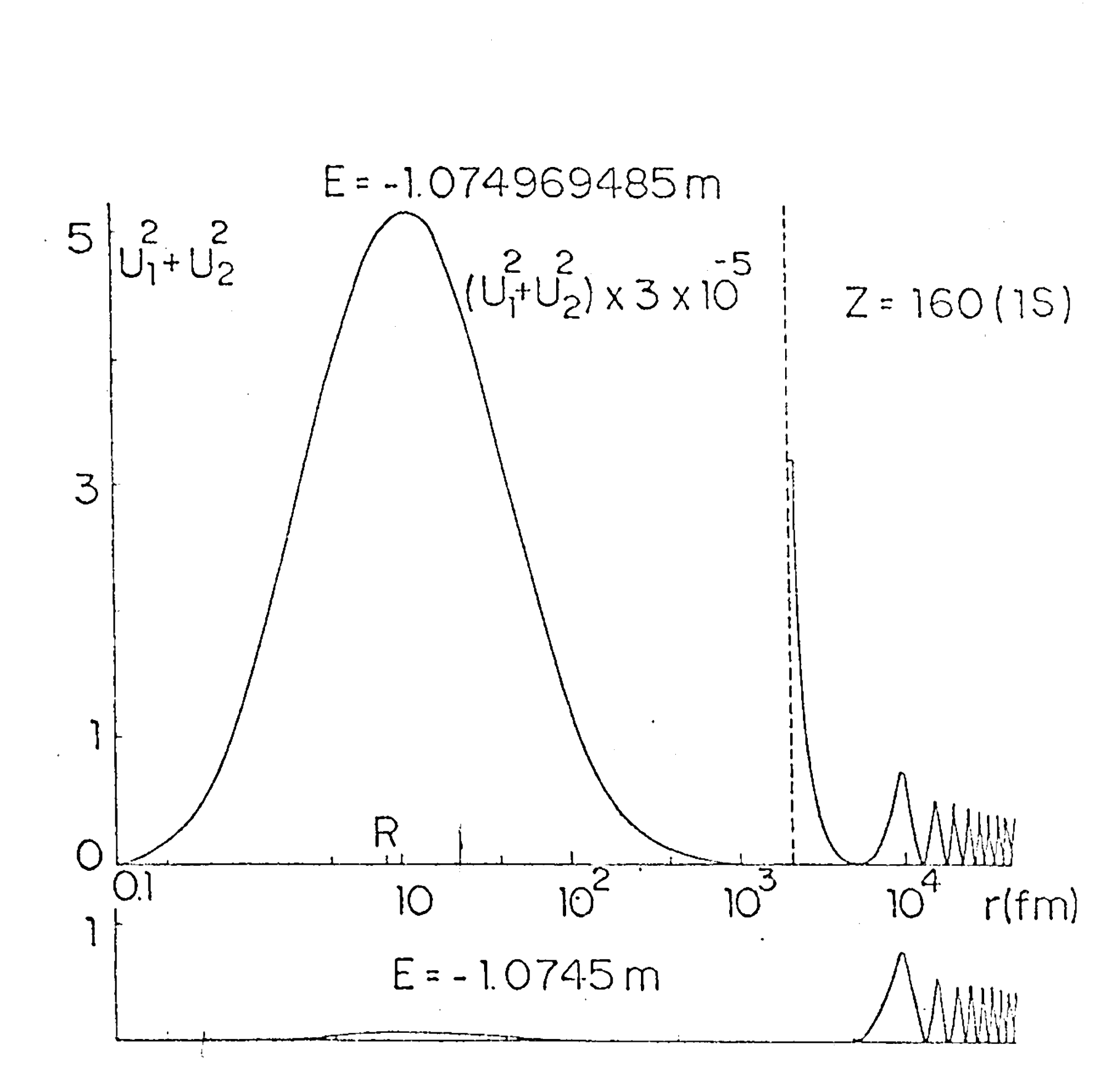}
\caption{Super-critical $U_1^2 + U_2^2$ function in the point nucleus $+$ AMDM model,
with $a = \alpha/2\pi$.}
\label{fig:2}
\end{figure}

\section{Conclusion and acknowledgments}

In conclusion, we have seen that the AMDM does regularize the wave functions at
$r = 0$, even for point nuclei.

As for the value of the critical atomic number, this will depend on the anomalous
magnetization the electron attains at strong fields. We also suggest that the combined
effects of both the finite size of the nucleus and the AMDM should be explored before a
conclusion as to the precise value of the critical charge can be reached.

We are deeply obliged to Prof.\ Walter Thomas Grandy, Jr.\ of the University of Wyoming
for suggesting the problem and for many useful discussions.

We want to thank the Brazilian agencies FINEP, CNPq and CAPES for partially sponsoring
this work.


\begin{thebibliography}{99}
\bibitem{1} V. S. Popov, Sov.\ Phys.\ JETP \textbf{32}, 526 (1971).
\bibitem{2} Y. B. Zeldovich and V. S. Popov, Sov.\ Phys.\ Usp.\ \textbf{14}, 673 (1972).
\bibitem{3} B. M\"uller, J. Rafelski and W. Greiner, Nuovo Cim.\ A \textbf{18}, 551 (1973).
\bibitem{4} J. Reinhardt and W. Greiner, Rep.\ Prog.\ Phys.\ \textbf{40}, 219 (1977).
\bibitem{5} J. Scheweppe et al., Phys.\ Rev.\ Lett.\ \textbf{51}, 2261 (1983).
\bibitem{6} W. J. Grandy, Private Communication.
\bibitem{7} J. J. Sakurai, \textit{Advanced Quantum Mechanics} (Addison-Wesley
Publishing Co., 1967).
\bibitem{8} A. O. Barut and J. Krauss, J. Math.\ Phys.\ \textbf{17}, 506 (1976).
\bibitem{9} M. Abramovitz and I. A. Stegun, \textit{Handbook of Mathematical Functions},
1st ed., p.~505 (Dover, New York, 1965).
\bibitem{10} Y. Kazama, C. N. Yang and A. S. Goldhaber, Phys.\ Rev.\ D \textbf{15},
2287 and 2300 (1977).
\bibitem{11} V. I. Ritus, Sov.\ Phys.\ JETP \textbf{48}, 788 (1978).
\bibitem{12} W. Pieper and W. Greiner, Z. Phys.\ \textbf{218}, 327 (1969).
\end{thebibliography}
\end{document}